\documentclass[]{aa}
\usepackage{times}
\usepackage{graphics}

\newcommand\om{$\Omega_0$}

\begin{document}
\thesaurus{}

\title{New light on the baryon fraction in galaxy clusters }

\author{R. Sadat\inst{1} \and A. Blanchard\inst{1,2}}
          
\institute{ 
Observatoire Midi-Pyr\'en\'ees, LAT, CNRS,14 Av. Edouard Belin, 31 400, Toulouse, France 
\and 
Observatoire astronomique de Strasbourg, ULP, CNRS, 
11, rue de l'Universit\'e, 67 000 Strasbourg, France 
	  }

\offprints{R. Sadat}
\mail{rsadat@ast.obs-mip.fr}
   
\date{Received \rule{2.0cm}{0.01cm} ; accepted \rule{2.0cm}{0.01cm} }

\authorrunning{Sadat \& Blanchard}
\maketitle
 
\begin{abstract} 

Baryon fraction in clusters, combined with constr\-aints from primordial nucleosynthesis is currently used to provide a robust upper limit on the cosmological density parameter $\Omega_0$. Current analyses lead to 
 gas fractions at virial radius which are typically of the order of 0.20$h^{-3/2}_{50}$ , favoring a low density universe.
In this work, we examine critically this issue through the analysis of the baryon distribution in clusters. We find that the currently derived gas fraction profile, increases regularly from the inner part to the outer part, up to the virial radius, and beyond. Such a shape contrasts with what is expected from numerical hydro-dynamical simulations, in which the gas fraction is more or less constant in the outer region, reaching a plateau when the contrast density falls off below $10^4$. 
We argue that  such a difference is hardly explained by reheating effects, while taking into account various factors entering into the determination of clusters gas fraction may erase such a difference. Indeed, using  recent estimates  on gas content in the outer part of clusters  (Vikhlinin et al., 1999) and applying the correction factor due to the effect of gas clumping, we find that the gas fraction shape over the range $ 200< \delta< 10^5$ is roughly consistent with hydro-dynamical simulations for an universal gas fraction in the range $8-11$\% (for $h_{50} =1.$),  the mass estimators being calibrated from numerical simulations. In contrast, values of the order of 20\% do not give acceptable fit to the data on any scale.
We conclude, that high values of $\Omega_0$ can not be ruled out on the basis of the baryon fraction argument. 
\end{abstract}

\section{Introduction}

One of the primary goal of cosmology is the determination of the mean density 
of the universe. Theoretical arguments such as the inflation predict a flat 
universe with $\Omega_0=1$. However, most of the observational data do not seem to 
provide a strong evidence for a critical universe.\\
 Rich clusters of galaxies are the largest virialized systems in the universe, they therefore provide us with useful constraint on \om. The evolution of the 
abundance of X-ray clusters of a given mass (temperature) is one of the most 
powerful test of $\Omega_0$ (Oukbir \& Blanchard 1992) that has been developed
 during these last years (Viana \& Liddle, 1996, Henry 1997, Eke et al. 1998, 
Sadat et al. 1998), but results are still contradictory (Blanchard et al. 2000). 
Clusters are the most massive objects
for which both the luminous baryonic mass (dominated by the X-ray emitting
intracluster gas) and the total gravitating 
mass can be determined and therefore an upper limit on the baryon fraction $f_b$ can be estimated. As
primordial nucleosynthesis can provide reliable constraints on the
value of the baryonic density parameter $\Omega_b \sim 0.03-0.08 h_{50}^{-2.}$
 ($h_{50}$ being hereafter the Hubble constant in unit of 50 km/s/Mpc; for numerical applications we used $h_{50}= 1.$), one can derive an upper limit to the matter density
parameter: $\Omega_0 = \Omega_b / f_b$. White et al. (1993), estimated the baryon fraction in Coma cluster of $f_b \sim 0.15 h_{50}^{-3/2}$ at the Abell radius and several recent analyses of large samples of clusters have confirmed such a high baryon fraction $f_{b}\approx 15 - 20\%$ (David et al, 1995, Evrard 1997, Mohr et al. 1999, Ettori \& Fabian 1999, EF99 hereafter). This fact suggests that the baryon content observed in rich clusters exceeds the universal value expected in an $\Omega_0=1$ universe. Since numerical simulations have shown that baryons cannot segregate significantly in clusters, this result implies that the mean density of the universe is lower than the critical value.

In this paper, we examine the issue of the baryon fraction in clusters, focussing on its distribution from both observational and numerical simulations sides.

\section{ The baryon fraction: Observations versus simulations.}
Let us first examine the main results obtained from theoretical side, mainly from numerical simulations, and compare them to observations.
The fate of gas in clusters is expected to be settled during the gravitational collapse of the mixture of baryons and dark matter. In the early phase of the collapse, both gas and dark matter are expected to follow the free fall trajectories until the gas is shock-heated which may lead to a segregation in the denser parts of clusters (corresponding more or less to the virialized region) between the gas and dark matter due to different physics governing the two fluids (Teyssier et al., 1997). In the case where only gravity is the source of heating of the gas, the baryon fraction is expected to follow a scaling law, depending  only on the density contrast (Kaiser 1986). 
Hydro-dynamical simulations help to study quantitatively such a behavior. 
As a fiducial model we use the results of the simulations developed by different groups (the Santa Barbara project) and nicely summarized by Frenk et al. (1999, F99 hereafter). The theoretical gas fraction distribution for a value of the global gas fraction of 16\% (corresponding to the range of what is generally inferred from the observations) is shown in Fig 1, as well as the range in which $2/3$ of the results from the simulations lie, taken as our 68\% confidence interval. 
On the observational side, the gas fraction at a given radius is inferred 
from the gas mass derived from the X-ray gas emissivity (see section 4) and the gravitating mass which 
can be determined from either the hydrostatic $\beta$-model or from the 
universal dark matter profile (Navarro et al., 1995, NFW) normalized to the temperature-mass relation 
calibrated from numerical simulations. Isothermal hydrostatic mass estimates are systematically lower than numerical simulations calibrated masses: Roussel et al. 
(2000, hereafter RSB00) found gas fractions 20\% smaller when virial masses are estimated from scaling laws calibrated from numerical simulations by Evrard, Metzler and Navarro, (1996, EMN96 hereafter). 
Taking into account for the temperature profile, hydrostatic equation would lead to masses nearly twice smaller (Nevalainen et al. 1999). Moreover, Balland and Blanchard (1997) have shown that even when accurate measurements of the temperature profile are available, hydrostatic mass estimates could suffer from a large uncertainty as later confirmed by Hughes (1998). To ensure consistency in the comparison with the simulations we restrict ourselves to the mass-temperature 
relations from numerical simulations and use the following two normalizations : $T_X = 4.75 (M_{200}/10^{15}M_{\odot})^{2/3} \textnormal{keV}$ (EMN96) and $T_X = 3.81 (M_{178}/10^{15}M_{\odot})^{2/3}$ \textnormal{keV} (Bryan and Norman 1998, BN98), $M_{\delta}$ being the mass enclosed in a region with a density equal to 
 $(1+\delta)$ times the critical density. These two values for the normalization can be considered to be the extreme values among existing numerical simulations, 
BN98  normalization leading to virial masses nearly 40\% higher.\\
We use the gas fractions estimated at various $M_{\delta}$ values from a compilation of X-ray data of a large sample of groups and galaxy clusters which are published in 
RSB00. For comparison, we also use the gas fractions at $r_{500}$ and $r_{200}$ (corresponding to $\delta=500$ and 200 respectively) quoted by Arnaud \& Evrard (1999, AE99) who used a similar approach to determine the virial mass. The resulting gas fraction distribution is shown in Fig. 1. We have also included the point at $r_{500}$ given by EF99; while 
their original value was 
17\% (based on hydrostatic mass estimates), we found that this value translates to  $f_g \sim 13.7$ at a radius of contrast density $\delta \sim 620$, when one is using EMN96 mass estimator. \\
Two striking features can be immediately noticed when we compare the gas fraction distribution derived from observations to the simulations.
  Firstly, in the central regions where the X-ray data are of better quality, a strong disagreement exists between the data and numerical predictions for primordial fraction of the order of $16\%$. Secondly, the gas fraction profile inferred from observations shows a strong trend of increasing from the inner parts to the outskirts, as also found by David et al. (1995), contrary to theoretical predictions which profiles flatten to approach more or less the asymptotic value close to the universal value.
Clearly, this discrepancy calls for caution when one is using the baryon (gas) fraction to set upper limit on the mean density of the universe.

\begin{figure}[htbp]
\resizebox{9cm}{!}{\includegraphics{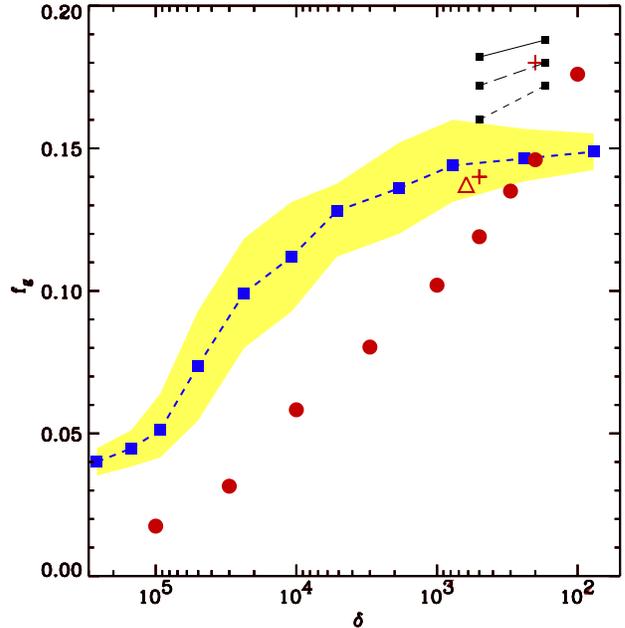}}
\hfill
\parbox[b]{87.5mm} {\caption{The distribution of the gas fraction $f_g$ versus $\delta$ ($\delta \equiv \rho(<r)/\rho_c$ where 
$\rho_c \equiv 3H^{2}_{o}/8\pi G$). The observations are from RSB00 
 {\it filled circles}, EF99 {\it empty triangle} and AE99 {\it crosses}. Statistical uncertainties on these quantities are smaller than the symbols size. The {\it squares} connected by a 
dashed line correspond to the theoretical gas fraction for a global value of $16\%$. {\it Small squares} connected by a line correspond to the gas fraction calculated from 
numerical simulations including winds by Metzler \& Evrard (1997): for a 6 keV cluster ({\it long dashes}), a 3 keV cluster ({\it small dashes}) and without including winds ({\it continuous line}) with a global value of 20\%.}
 \label{Fig:fig1.ps }}
\end{figure}
 
\section{Is the baryon discrepancy robust?}
Let us now examine what could be  the origin of this discrepancy. A first possibility is that the numerical simulations are suffering from numerical limitations that 
forbid a correct description of the baryon fraction shape. A second possibility is that the physics involved in the modeling of clusters is not adequate.
A third possibility is that 
the strengh of the conclusions is not warranted by the data. In this section 
we will examine the first two possibilities and comment in the next section on the third one.\\
Numerical simulations have been extensively used in the modern cosmology. Several groups in the world have designed hydrodynamic numerical codes based on different techniques to simulate the gas dynamics in the cosmological context. 
F99 provides a comparison between these different codes allowing to test the ability of numerical simulations to reproduce the characteristics of cosmological formation of a cluster, assuming no cooling and heating processes. This comparison has been performed on the results of simulated X-ray clusters. F99 found that all the simulations agree to better than 10\% on the gas mass at the virial radius and that the mean behavior of the gas fraction appears quite similar in all the simulations. Although  noticeable dispersion in the absolute value of the gas fraction exists, the general behavior is always similar:  the gas fraction raises up in the central part and flattens in the outer part. This suggests that numerical simulations are reliable on this issue given the assumed physics and thus the origin of the above difference cannot be due to numerical problems.

A second possibility, which seems more likely, lies in the  physics of the formation of clusters: most of the numerical simulations assume adiabatic gas physics while there are convincing evidences that non gravitational processes are playing a significant role, otherwise the observed $L-T$ relation could not be recovered. Heating of the IGM before or after the formation process is the often advocated mechanism (Cavaliere et al. 1998). 
The consequence of heating is to affect the gas distribution inside the cluster and consequently the radial distribution of $f_{b}$. One may think that this effect could provide a natural explanation to the steep increase of $f_b$ observed in the data. However, we believe that there are two reasons why this is not the most probable explanation (of course there is enormous freedom in designing a reheating scenario, so excluding totally this possibility is not possible) : {\it i}) even when extreme models with heating are considered, the asymptotic slope of the gas fraction is not as steep as observed. Fig. 1, shows the theoretical gas fraction at 
$\delta=500$ and 200 resulting from the simulations which include energy injection 
by SNe-driven winds (Metzler \& Evrard, 1997). As we can see, the effect on the asymptotic slope is not very important. Even in the extreme cases of heating as illustrated by the small dashed curve in Fig. 1 (corresponding to low temperature clusters for which heating is expected to be very efficient), the predicted shape does not show the steep rising of the observed distribution of the gas fraction; {\it ii}) in the presence of strong heating large departure from the scaling laws is expected : the gas fraction as well as the asymptotic slope of the gas profile are expected to vary significantly with temperature. Although there is some debate on these questions, several groups have found no clear evidence of strong variations of the gas fraction with temperature (RSB00, Mohr et al. 1999, Vikhlinin et al. 1999, VFJ99 hereafter). Moreover, it has been shown that the emission from the gas follows roughly the scaling law in the outer regions (Neumann \& Arnaud, 1999, VFJ99, RSB00) for clusters with $T \approx 1-10$ keV. Despite the wide range of possibilities on heating mechanism, and the limited number of cases that have been investigated, it seems to us difficult to significantly change the asymptotic slope and preserve at the same time the scaling of the gas fraction over a rather wide range of contrast densities.\\
These two arguments provide serious evidence that the apparent  rising of the gas distribution in the outer parts of clusters is unlikely to be the result of some heating process.

\section{Some remarks about gas estimation.}

In this section  issues concerning the reliability of the gas fraction as estimated from the observations are discussed. 

\subsection{From observations to gas masses}

The X-ray gas masses are usually derived by fitting the X-ray surface brightness $S(r)$ to a $\beta$-model,  $S_{o}(1+(\theta/\theta_{c})^2)^{-3\beta+1/2}$ where $\theta$ is the projected angular distance to the center. The mean value of the slope $\beta$ is generally found to be around 0.65 smaller than the theoretical asymptotic value derived from the simulations $<\beta>=0.8 - 1$. 
However, the X-ray gas emission is rarely traced out to very large radii, so that gas fractions are measured only up to an X-ray limiting radius $R_{Xlim}$ for which the signal-to-noise is good enough. Most of  the time, it was necessary  to extrapolate to the virial radius $r_{200}$ or to some other outer radius like $r_{500}$. However, it is essential to realize by doing so, that gas masses are estimated at a radius at which the actual emission is not well known, poorly constrained and with a value of the order of the X-ray background or less. Furthermore, the emissivity being dominated by the central region, a parametric fit will be rather insensitive to the outer profile. The extrapolation of a typical $\beta$-model to asymptotic regions ($r \approx r_{v}$) leads to $\rho_{gas} \approx r^{-3{\beta}} \sim r^{-1.95}$ (for $<\beta>\sim 0.65$) which is significantly less steep than the universal dark matter profile  (NFW) found in the simulations $\rho_{DM} \approx r^{-2.7}$. These two facts may actually explain the trend of steep increasing in the $f_b$ profile (Fig. 1). 
It is therefore vital to examine whether the extrapolation of the $\beta$-model up to the virial region does or not introduce a bias in the gas estimations. Recently, a large sample of X-ray ROSAT images in which the data trace the emission up to very large radii have been analyzed by VFJ99.  
They found that the $\beta$-model does not provide an accurate description of the surface brightness over the whole range of radii, the slope in the outer regions (0.3$r_{200}$ to $r_{200}$) being actually steeper than with the one found when the $\beta$-model is applied to the entire cluster. 
This result is very important as it shows that a dedicated treatment of the outer parts of the cluster leads to a different gas distribution than what is generally assumed and provides direct evidence that the extrapolation to the virial radius of the  $\beta$-model could be unsafe. \\ 

\subsection{The clumping of the gas}

Most of the previous studies of the gas and mass distribution assume that clusters are spherically symmetric and that the ICM is uniform. In reality, most of the clusters exhibit a certain level of density fluctuations at small and large scales due to accretion and merger events which continuously occur inside clusters. This clumping may introduce a further potential bias in gas mass estimates and therefore in the inferred gas fractions. Furthermore it is reasonable to assume that the clumping of the gas is more pronounced in the outer part, as the gas may  not have enough time to relax. 
If the gas is locally inhomogeneous, it is useful to introduce the clumping factor $C = <\rho_g^2>/<\rho_g>^2$ where $\rho_g$ is the gas density. In presence of clumping, gas mass estimates are overestimated by roughly $C^{1/2}$. 
Moreover, this factor is likely to increase in the outer parts.  
Studying the effects of density fluctuations on small scale and large scale 
substructures on the ICM mass measurement of a set of simulated clusters, 
Mathiesen et al. (1999, MEM99) have quantified the clumping effect and found that the estimation of the gas mass from models assuming a uniform density carries 
a bias of $\sqrt{C} \sim 1.16$ at $\delta = 500$. It is worth 
noticing that this factor is probably still poorly known, as its measured  
value might be affected by the numerical simulation limitation. Interestingly enough is to notice that although the various simulations reported in the 
Santa Barbara project do find a nearly constant baryon profile in the outer 
regions of the cluster, much larger dispersions are found in the emissivity at large radii
 (more than a factor of 2, see their figure 19), which may either reflect the numerical limitations or the differences in the clumping factor treatment.
 
\section{The baryon fraction revisited}

Here we evaluate quantitatively the consequence of the above biases. From what we have seen, robust gas mass estimates require that the emissivity is actually detected up to the radius at which gas masses are given (notice that this may still introduce a kind of Malmquist bias). To our knowledge, only VFJ99 sample
satisfies this criteria: their published results allow us to compute the gas mass content enclosed within two different radii, $R_{1000}$ and $ R_{2000}$ (see their equation 2):
$M_g(\delta) = 4{\pi}/3(1+\delta)\rho_b R_{\delta}^3$ (using their $\rho_b$).
We then computed the mass enclosed within the same physical radius assuming 
an universal profile (NFW) with a concentration parameter $c = 5$. As they found a slight temperature dependence we work at their median temperature $T = 6$ ${\rm keV}$. We obtain $f_g (\delta = 770) = 0.1064$ and $f_g (\delta = 343) = 0.1194$ with EMN calibration and $f_g (\delta = 974) = 0.0842$ and $f_g (\delta = 437) = 0.0934$ with BN98 normalization. These values are similar to those derived by RSB00, and smaller than those of AE and EF by nearly 2\%.\\
We then correct for the clumping effect of the gas: not much is known about its properties, we therefore restrict ourselves to correct only the outer value at $\delta = 500$, using the factor given by MEM99. In the inner parts of the cluster we use RSB00 gas fractions, assuming that no correction has to be applied to them. The resulting values are shown as big stars and circles in Fig. 2.
By matching the point at $\delta =500$, we  obtain  the following primordial gas fractions : $f_g = 0.0875 \pm 0.0075 \: (68\%)$ (BN98) and  $f_g = 0.108 \pm 0.0092$ (EMN96), the uncertainties being due to numerical simulations.
Two important features arise:   {\it i}) The  gas fraction  now appears to be consistent over a large range of $\delta$ with numerical simulations predictions: in particular we can see that the asymptotic flattening is now recovered. Note that, the observations fall off slightly below the predicted profile in the inner parts ($\delta >10^{4}$), this could well be due to energy injection, {\it ii}) the inferred gas fraction at $r_{500}$ is now lower than what is previously found (Fig. 1). The uncertainty being due to the uncertainty in the calibration of the $M-T$ relation from numerical simulation.\\
\begin{figure}
\resizebox{9cm}{!}{\includegraphics{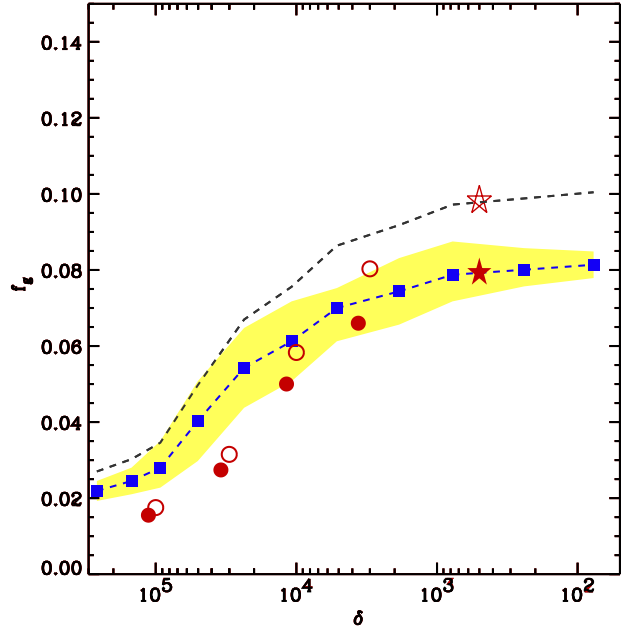}}
\hfill
\parbox[b]{87.5mm} {\caption{The distribution of the gas fraction $f_g$ versus $\delta$: {\it circles} correspond to the observed $f_g$ from RSB00 but restricted to the region where the gas is detected (no extrapolation), using EMN96 ({\it empty circles}) and BN98 ({\it filled circles}) calibration. {\it  Empty} ({\it filled}) {\it star} corresponds to the estimated $f_g$ using VFJ99 sample and corrected for the clumping with EMN96 (BN98) calibration. {\it Filled squares} linked by a dashed line correspond to the gas fraction predicted by numerical simulations for a global fraction of 8.75\%. The upper dashed line corresponds to the case where the global fraction is 10.8\% }
 \label{Fig:fig2 }}

\end{figure}
  In order to estimate the total baryon fraction we add the stellar contribution, taking a mean value of the order of $1\%$ (RSB00). 
Our final values  are  : $f_{b} \sim 0.10 \pm 0.01$ (with BN98 calibration) and $f_{b} \sim 0.12 \pm 0.01$ (with EMN96 calibration).\\
It is worth noticing that the baryon fraction shape inferred in this present work (Fig. 2) corresponds to a steeper gas distribution asymptotic slope than the one derived from the standard $\beta$-model and even steeper than VFJ99 slopes. Plugging this gas profile into the isothermal hydrostatic equilibrium would lead to masses which are much higher than previous estimates and thus more consistent with masses calibrated from numerical simulations.

\section{Implications on $\Omega_0$}
The arguments presented above clearly show that previous mean baryon fraction estimations at the virial radius are probably biased due to the extrapolation of data and to the effect of gas clumping. 
This leads us to re-examine the constraint on $\Omega_0$ set by the baryon fraction. Following the notation of White et al. (1993) :
\begin{equation}
\Omega_0 = \gamma(r_{500})\frac{\Omega_{b}}{f_{b}(r_{500})}
\end{equation}
where $\gamma(r_{500}) \sim 0.91$ is the baryon depletion as predicted by the 
simulations. $\Omega_{b}$ is set by the primordial nucleosynthesis theory through the  
baryon-to-photon parameter $\eta_{10}$ which is constrained by the measurement 
of the abundance of light elements. The main uncertainty comes from the deuterium abundance (D/H) measurement. There are two conflicting estimates 
of D/H, a high value which leads to $\eta_{10} \sim 2$ (Rugers \& Hogan 1996) 
and a low value corresponding to $\eta_{10} \sim 5.6$ (
 O'Meara et al.,  2000). This latter
 value is consistent with that measured by Wadsley et al. (1999) using the He 
Lyman--$\alpha$ forest and the amount of baryons in order to produce the 
observed  Lyman--$\alpha$ systems (Rauch et al., 1997; Weinberg et al., 1997)
 and are favored by CMB fluctuations properties (Le Dour et al, 2000; Tegmark \& Zaldarriaga, 2000). Taking $ \Omega_b h_{50}^2 \sim 0.08$ (assuming that all baryons in clusters are seen) and using our estimated primordial baryon fraction obtained with BN98 calibration ($f_{b}= 0.10\pm 0.01$), we estimate the density of the universe: $\Omega_0 = 0.8 \pm 0.1 $ in agreement with the value derived from the evolution of the X-ray clusters abundance (Blanchard et al. 2000).\\
If we use the primordial baryon fraction obtained with EMN96 calibration we find $\Omega_0 \sim 0.65$.
  
\section{Conclusion }
Baryon fractions derived from X-ray 
clusters of galaxies are generally in the range (15\% - 20\%) (for $h_{50} =1.$). This 
 exceeds the value predicted by Big Bang nucleosynthesis in an $\Omega_0 = 1$ 
universe which has led to the conclusion that we are living in a low-density 
universe with $\Omega_0 \sim 0.2 - 0.4$. However, we found that apparent gas fraction profiles, which obey reasonably well to scaling laws, differ in their shape from 
what is found in  numerical simulations: these
gas fraction profiles seem to continuously increase from the inner to the outer parts, while it flattens asymptotically, approaching the universal value at the 
virial radius in the numerical simulations. \\
Using updated X-ray data on the gas distribution out to large radii 
from VFJ99, binding mass based on numerical simulations and correcting for the clumping effect, we obtain the following results: {\it i}) the overall shape of the baryon fraction is roughly consistent with numerical simulations. {\it ii}) This agreement is obtained provided that the global baryon 
fraction is of the order of  10\%, while higher values of the order of 15--20\% are clearly ruled out. {\it iii}) This revised value of the baryon fraction, combined with current limits on $\Omega_{b}$ from primordial nucleosynthesis, is consistent with a high density of the universe.
Thus we conclude that the apparent discrepancy between the observations and numerical simulations in the baryon fraction distribution inside clusters can be erased once the above systematics are taken into account. This work has also raised the following question: is the baryon fraction a strong argument against $\Omega_{0}=1$? Several effects and systematics may act in concert to create the apparent gap between the current estimates of the baryon fraction in clusters and the universal value, as it is shown in this work, therefore it seems to us that the baryon fraction is no longer a strong and ``clean'' argument in favor of a low density universe as is too oftenly claimed. It is therefore important to better constrain and control all the systematics acting in the gas fraction measurements, this will be soon possible with XMM, AXAF and SZ experiments. On the other hand numerical simulations of X-ray clusters are also needed to quantify the clumping effect as well as the role of non-gravitational physics.

\begin{acknowledgements} We would like to thank the anonymous referee for his useful remarks which have significantly improved this work. R.~Sadat acknowledges the support by the R. Chr\'etien International Research Grant. 
\end{acknowledgements}

\end{document}